\documentclass[cameraready]{Interspeech}      % [CR] camera-ready mode
\usepackage[T1]{fontenc}
\usepackage[utf8]{inputenc}
\title{Word Lengthening as a Function of Utterance Position: A Multi-Corpus Study}

\author[affiliation={1,2}, correspondingauthor]{Mateo}{C\'{a}mara}
\author[affiliation={1}]{Jos\'{e} Luis}{Blanco}
\author[affiliation={3}]{Juan Ignacio}{Godino}
\author[affiliation={2}]{Jeung-Yoon}{Choi}
\author[affiliation={2}]{Stefanie}{Shattuck-Hufnagel}

\address{
    $^1$ \blue{Signal Processing Applications Group, Information Processing \& Telecomm. Center, Universidad Polit\'{e}cnica de Madrid, Spain} \\
    $^2$ Speech Communication Group, Research Laboratory of Electronics, Massachusetts Institute of Technology, USA \\
    $^3$ Bioengineering and Optoelectronics Lab., Universidad Polit\'{e}cnica de Madrid, Spain
}

\email{\{mateo.camara, jl.blanco, ignacio.godino\}@upm.es, \{jyechoi, sshuf\}@mit.edu}


\newcommand{\blue}[1]{#1}   % [CR] finalized: blue draft markup flipped to black

\usepackage{comment}
\usepackage{graphicx}

\begin{document}
\maketitle

\begin{abstract}
Efficient turn-taking requires interlocutors to predict turn endings \blue{within a few hundred milliseconds}. Beyond syntactic and pragmatic completion, prosody (especially pre-boundary lengthening) supports projection. We test whether turn-final words are longer than mid-sentence words, whether this reflects prosodic modification rather than lexical choice, and where within the word it concentrates. We analyze four corpora spanning styles and two languages (English, Spanish): Switchboard, Columbia Games, BU Radio, and Glissando, with >500 speakers, $39{,}470$ turn-final and $206{,}268$ mid-sentence tokens across $\sim39{,}500$ turns. Turn-final words are longer (mean \blue{${\approx}191$\,ms}; $d=1.14$). The effect persists in matched-word, within-speaker comparisons (\blue{$80$\,ms}; $p<0.001$) and is localized mainly to the final syllable ($d=0.89$). Turn-final lengthening thus emerges as a robust, localized cue to floor transfer.
\end{abstract}

\section{Introduction}
Human conversation is characterized by \blue{close} temporal synchrony. The gap between one speaker stopping and another starting is often on the order of 200\,ms, shorter than typical speech-production latencies \cite{Sacks1974,LevinsonTorreira2015}. This \blue{suggests that listeners anticipate likely completion points and begin planning in advance, rather than pinpointing the exact moment of completion}, using cues from syntax, semantics, and prosody as the current turn unfolds \cite{HeldnerEdlund2010, deRuiter2006}.

While syntax and semantics provide structural constraints, prosody offers immediate acoustic cues. One such cue is \textit{lengthening} near the end of prosodic units. The increase in segment durations near prosodic boundaries is a well-documented phenomenon in read and laboratory speech \cite{Klatt1975,Wightman1992,TurkShattuck2007,White2014}. From a mechanistic perspective, boundary-related lengthening appears because articulatory movements gradually slow as speakers approach the end of a prosodic unit \cite{ByrdSaltzman2003}. Prosodic boundaries may also be marked by stronger, more clearly defined articulatory gestures \cite{FougeronKeating1997}. At the same time, the magnitude of lengthening depends on boundary strength (as encoded, for example, in ToBI break indices \cite{BeckmanAyersElam1997} that distinguish phrase-internal, intermediate, and intonational-phrase boundaries). It motivates graded analyses rather than a single ``final vs.\ non-final'' contrast \cite{KentnerEtAl2023}. \blue{Although phrase-final lengthening has been studied most extensively in read speech, utterance-final lengthening has also been documented cross-linguistically in spontaneous speech \cite{SeifartEtAl2021}.} Turn boundaries in conversation constitute higher-order prosodic and interactional boundaries where similar timing mechanisms may operate. We do not claim novelty for final lengthening, but quantify position-dependent duration differences in a turn-taking framework across corpora and styles (a single acoustic-prosodic cue, not a turn-end predictor). We investigate whether turn-final lengthening generalizes across corpora and languages, and if the durational differences reflect prosodic modification or simply lexical selection.

This paper studies turn-related lengthening across four corpora spanning spontaneous, task-oriented, and read speech, and English and Spanish: Switchboard \cite{LDC97S62}, Columbia Games \cite{LDC2021S02}, BU Radio News \cite{LDC96S36}, and Glissando \cite{GarridoEtAl2013}. We address three questions:
\begin{enumerate}
    \item Is turn-final lengthening a robust phenomenon across different speech styles (spontaneous vs.\ task-oriented vs.\ read) and languages (English vs.\ Spanish)?
    \item Is the observed lengthening an artifact of lexical choice (e.g., using longer words at the end of sentences) or a true prosodic modification?
    \item Is the lengthening distributed across the word or localized to the final syllable?
\end{enumerate}

% [CR] Removed the section-roadmap sentence to save space; it was low-value and its "Sec. II--VII" numbering was inconsistent with the rendered (arabic, eight-section) numbering. Original preserved:
% The remainder of the paper is organized as follows: Sec.~II reviews related work, Sec.~III describes the corpora and labeling, Sec.~IV details the measures, Sec.~V presents the main results, Sec.~VI reports robustness analyses, and Sec.~VII concludes.

\section{Related Work}
We organize prior work by linguistic level, from conversational turns down to syllables. We focus on how temporal cues support turn projection and turn transition timing.

\subsection{Turn-level} Conversation Analysis characterizes turn-taking as a coordinated system in which next speakers routinely join with minimal delay. It implies that completion is treated as \emph{projectable} (its upcoming end can be anticipated from cues in the turn) \cite{Sacks1974,ochs1996interaction}. Quantitative work corroborates this by showing that gaps are typically very short (often on the order of ${\sim}200$\,ms) relative to production latencies. This is consistent with listeners preparing responses before the current turn ends \cite{LevinsonTorreira2015}. De Ruiter \textit{et al.}\ argue that this projection is supported by the incremental convergence of partial cues (pragmatic action completion, syntactic closure, prosodic completion) rather than a single decisive trigger \cite{deRuiter2006}. Timing analyses of pauses, gaps, and overlaps further show that conversational ``synchrony'' follows structured distributions, constraining what counts as a smooth handover \cite{HeldnerEdlund2010}.

\subsection{Phrase-level} At the prosodic-phrase edge, pre-boundary lengthening is one of the most robust temporal correlates \cite{Klatt1975,Wightman1992,TurkShattuck2007}. In read speech, Klatt reported that phrase-final vowels were longer by about 40\,ms on average than comparable non-final vowels \cite{Klatt1975}. \blue{Beyond read speech, Seifart \textit{et al.}\ documented utterance-final word lengthening in spontaneous speech across ten typologically diverse languages, establishing its cross-linguistic robustness \cite{SeifartEtAl2021}.} Byrd and Saltzman's $\pi$-gesture model formalizes lengthening as a boundary-related gesture that competes with ongoing articulation, naturally predicting graded effects \cite{ByrdSaltzman2003}. Prosodic annotation frameworks operationalize boundary strength and enable systematic links between boundary labels and timing outcomes \cite{BeckmanAyersElam1997,KentnerEtAl2023}.

\subsection{Word-level} Word edges provide a practical point where prosodic phrasing and turn management meet. In task-oriented dialogue, Gravano and Hirschberg showed that turn transitions correlate with a bundle of acoustic-prosodic cues, with final-word duration among the informative predictors \cite{GravanoHirschberg2011}. Local \textit{et al.}\ described phonetic configurations (including voice quality and laryngeal adjustments around silences) associated with turn-holding versus turn-yielding outcomes \cite{Local1986}. Local and Walker further demonstrated that fine phonetic detail at word endings supports listeners' projection of completion \cite{LocalWalker2012}.

\subsection{Syllable-level} Syllable-based analyses help localize where temporal expansion is realized and distinguish targeted boundary marking from diffuse slowing. Oller reported sizeable final-syllable increases in controlled materials, with vowel durations increasing by roughly 100\,ms on average \cite{Oller1973}. Subsequent work suggests that boundary-related lengthening is typically strongest in the final syllable's rhyme \cite{Wightman1992,TurkShattuck2007,White2014}, and that its magnitude scales with boundary strength \cite{KentnerEtAl2023}.

\section{Data and labeling}

\subsection{Corpora}
We use four corpora spanning conversational and read speech: Switchboard (telephone conversation) \cite{LDC97S62}, Columbia Games (task-oriented dyadic dialogue) \cite{LDC2021S02}, BU Radio News (read broadcast speech) \cite{LDC96S36}, and Glissando (Spanish/Castilian speech) \cite{GarridoEtAl2013}. All analyses use corpus-provided word boundaries. These corpora \blue{were selected to maximize diversity in speech style and language (English and Spanish), while varying in the prosodic annotations they provide}. All are established resources with validated annotations.

\subsection{Turn position} Each token is labeled as \textbf{turn-final} (TF) if it is the last word before a speaker change or a long silence marking turn completion and \textbf{mid-sentence} (MS) otherwise. Note that sentence endings occurring mid-turn (where the same speaker continues) are classified as mid-sentence, not turn-final. This distinction is critical because syntactic completion does not always coincide with turn completion. This yields $39{,}470$ turn-final and $206{,}268$ mid-sentence tokens across all four corpora (Columbia Games: $7{,}437$/$66{,}417$; Switchboard: $17{,}151$/$79{,}736$; BU Radio: $7{,}821$/$23{,}954$; Glissando: $7{,}061$/$36{,}161$).

\subsection{Backchannels} Short listener responses (e.g., ``yeah'', ``okay'', ``mm-hm'') were identified using corpus-specific annotations and a unified word-list. This excluded $14{,}525$ tokens ($5.9\%$). The baseline analysis excludes these tokens.

\subsection{Prosodic boundary strength (break index)}
Corpus-provided ToBI-style break indices are available for Switchboard and BU Radio. Break-index labels were aligned to word tokens using file identifiers and word time spans. We group indices into \textbf{low} (0--1; within-phrase), \textbf{medium} (2; intermediate phrase), and \textbf{high} (3--4; intonational/major boundary) following \cite{BeckmanAyersElam1997}.

\subsection{Syllable structure}
Syllable durations are analyzed in corpora with time-aligned segmentation (Switchboard, Glissando). Final vs.\ non-final comparisons use polysyllabic words only.

\section{Measures and statistical reporting}
We report mean durations, differences in milliseconds, ratios, and Cohen's $d$. For two independent groups with means $\mu_1,\mu_2$ and standard deviations $s_1,s_2$ (sizes $n_1,n_2$), we use
\[
d=\frac{\mu_1-\mu_2}{s_p},\quad
s_p=\sqrt{\frac{(n_1-1)s_1^2+(n_2-1)s_2^2}{n_1+n_2-2}}.
\]
All group comparisons use Welch's $t$-test. For matched-word analyses, we pair the same word from the same speaker in both positions (\emph{strict}: same session; \emph{relaxed}: any session) and test paired differences with paired $t$-tests.

\section{Results}
\subsection{Turn-final words are significantly elongated}
\begin{figure}[t]
\includegraphics[width=\linewidth]{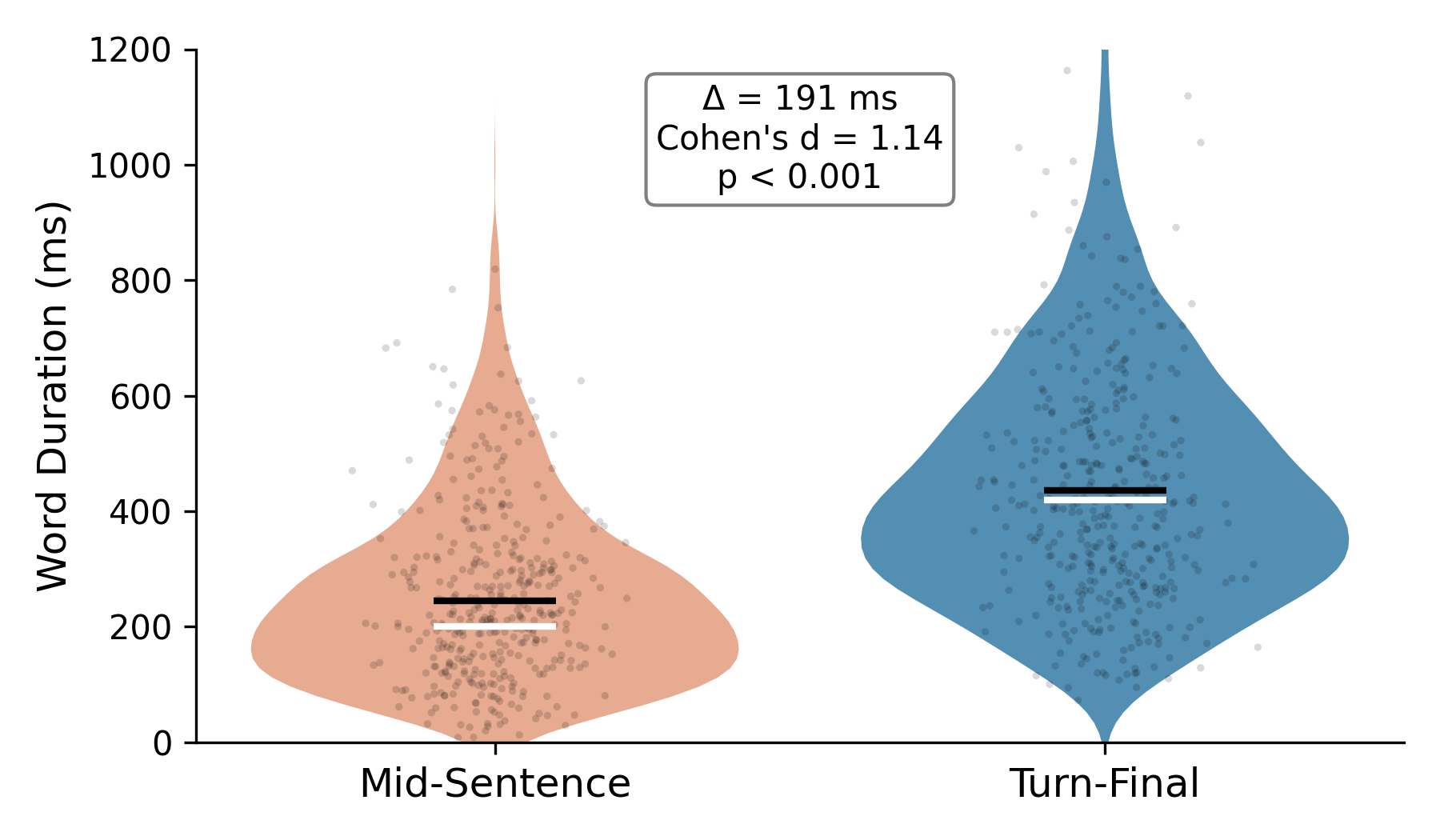}
\caption{Word duration over all datasets (pooled) based on sentence position.}
\label{fig:main_corpus}
\end{figure}

In the baseline setting (backchannels excluded), turn-final words have a mean duration of $0.44$\,s compared to $0.24$\,s for mid-sentence words (Fig.~\ref{fig:main_corpus}), yielding a difference of $203.1$\,ms (ratio $=1.85\times$) with a large standardized effect ($d=1.22$).

Table~\ref{tab:corpus} reports corpus-wise means and effect sizes. Effect sizes range from $d=0.78$ (Columbia Games) to $d=1.47$ (Glissando), with the direction consistent across all four corpora despite 4--9$\times$ sample imbalance. Absolute durations are higher in the read-speech corpora, reflecting their slower speaking rate.
% [CR] Removed the read-speech cross-linguistic comparison here because it duplicated Sec. 5.5 (Cross-linguistic comparison); that subsection is now its sole home. Original sentence preserved for reference:
% Comparing the two read-speech corpora to reduce style confounds: BU Radio (English, $\Delta=201$\,ms, $d=1.08$) and Glissando (Spanish, $\Delta=254$\,ms, $d=1.47$) both show clear turn-final lengthening, with a stronger effect in Spanish.

% recuerda: en el preámbulo
% \usepackage{graphicx}

\begin{table}[t]
\centering
\caption{Turn-position effect by corpus (backchannels excluded). $N_{TF}$/$N_{MS}$: number of turn-final and mid-sentence tokens. TF/MS: mean word duration in seconds. $\Delta$: difference in milliseconds. $d$: Cohen's $d$ effect size.}
\label{tab:corpus}
\setlength{\tabcolsep}{3.2pt}

\resizebox{\columnwidth}{!}{%
\begin{tabular}{l r r r r r r}
\toprule
Corpus & $N_{TF}$ & $N_{MS}$ & TF (s) & MS (s) & $\Delta$ (ms) & $d$ \\
\midrule
Columbia Games & 7{,}437  & 66{,}417 & 0.38 & 0.25 & 132 & 0.78 \\
Switchboard    & 17{,}151 & 79{,}736 & 0.40 & 0.22 & 179 & 1.19 \\
BU Radio       & 7{,}821  & 23{,}954 & 0.50 & 0.30 & 201 & 1.08 \\
Glissando      & 7{,}061  & 36{,}161 & 0.51 & 0.26 & 254 & 1.47 \\
\bottomrule
\end{tabular}%
}
\end{table}

\subsection{Matched words reveal persistent lengthening effect}
To control for lexical composition, we compare durations of the \emph{same word} produced by the \emph{same speaker} in turn-final and mid-sentence contexts. We construct matched pairs by indexing tokens by \texttt{(speaker, word)} and selecting one turn-final and one mid-sentence occurrence per pair\blue{, rather than enumerating all pairwise permutations, to avoid overweighting frequent words}; in the \emph{strict} setting, we additionally require both occurrences to come from the same recording/session.

\begin{figure}[t]
\centering
\includegraphics[width=\linewidth]{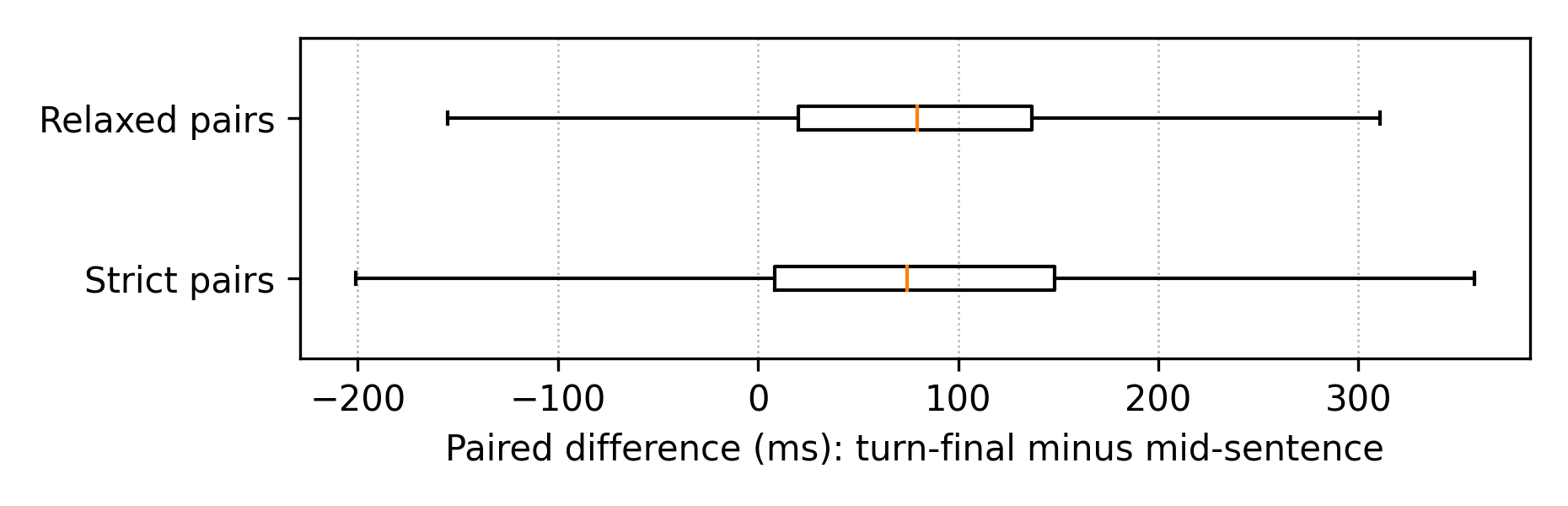}
\caption{Distribution of paired duration differences (ms): turn-final minus mid-sentence, for strict and relaxed matched-word pairs (backchannels excluded).}
\label{fig:matched_pairs_boxplot}
\end{figure}

Across \textbf{strict} pairs ($N=9{,}018$), turn-final tokens are longer by $79.9$\,ms on average (median $74.5$\,ms; paired $d=0.59$; $77.1\%$ positive); across \textbf{relaxed} pairs ($N=6{,}547$), by $81.1$\,ms (median $79.5$\,ms; $d=0.69$; $81.6\%$ positive). Fig.~\ref{fig:matched_pairs_boxplot} shows the distribution of paired differences.

A minority of pairs reverse ($22.9\%$ strict, $18.4\%$ relaxed). Per-word, only 32 of 484 words ($6.6\%$) are negative, concentrated among backchannel-like items with clipped turn-final delivery (e.g., ``yep'', ``nope'', ``mm''). Overall, $92.8\%$ of words show positive turn-final effects.

\subsection{Word-final syllables carry most of the effect}
% [CR] The original paragraph here was misplaced break-index content duplicating Sec. 5.4 (Prosodic boundary strength). It has been replaced with the syllable-localization description that matches this subsection header and Table 2. Original text preserved below for reference:
% If turn-final lengthening reflects boundary-related timing control, word duration should also track independently annotated prosodic boundary strength. Using ToBI-style break indices (Switchboard and BU Radio; Sec.~III-B), Fig.~\ref{fig:breakindex} shows a clear increase with boundary category: low-boundary words average $0.22$\,s, medium $0.34$\,s, and high $0.44$\,s ($\Delta=214.8$\,ms; $d=1.35$ for high vs.\ low), supporting a prosodic-boundary interpretation \cite{Wightman1992,TurkShattuck2007}.
\blue{Using the corpora with time-aligned syllable segmentation (Switchboard and Glissando) and polysyllabic words only, Table~\ref{tab:syllable_localization} localizes the effect within the word. Within turn-final words, the final syllable is substantially longer than the non-final syllables ($0.29$ vs.\ $0.18$\,s; $d=0.89$), with a comparable within-word pattern for mid-sentence words ($0.27$ vs.\ $0.18$\,s; $d=0.77$). Comparing the same syllable position across turn positions, the turn-final vs.\ mid-sentence contrast is concentrated in the final syllable ($d=0.09$) and is essentially absent in non-final syllables ($d=0.01$). The turn-position effect is thus carried almost entirely by the word-final syllable, consistent with boundary-adjacent timing control \cite{ByrdSaltzman2003}.}

\begin{table}[t]
\centering
\caption{Syllable localization by turn position (polysyllabic words). Top: within-turn final vs.\ non-final. Bottom: across-turn turn-final vs.\ mid-sentence.}
\label{tab:syllable_localization}
\setlength{\tabcolsep}{3.5pt}

\resizebox{\columnwidth}{!}{%
\begin{tabular}{lrrrrr}
\toprule
\multicolumn{6}{l}{\textbf{Within-turn (Final vs.\ Non-final)}}\\
\midrule
Turn pos. & $N_F$ & Mean$_F$ (s) & $N_{NF}$ & Mean$_{NF}$ (s) & $d$ \\
\midrule
Turn-final   & 13{,}192 & 0.29 & 167{,}094 & 0.18 & 0.89 \\
Mid-sentence & 58{,}625 & 0.27 & 703{,}709 & 0.18 & 0.77 \\
\midrule
\multicolumn{6}{l}{\textbf{Across-turn (Turn-final vs.\ Mid-sentence)}}\\
\midrule
Syll pos. & $N_{TF}$ & Mean$_{TF}$ (s) & $N_{MS}$ & Mean$_{MS}$ (s) & $d$ \\
\midrule
Final     & 13{,}192  & 0.29 & 58{,}625  & 0.27 & 0.09 \\
Non-final & 167{,}094 & 0.18 & 703{,}709 & 0.18 & 0.01 \\
\bottomrule
\end{tabular}%
}
\end{table}

\subsection{Prosodic boundary strength mirrors turn-final lengthening}
\begin{figure}[t]
\centering
\includegraphics[width=\linewidth]{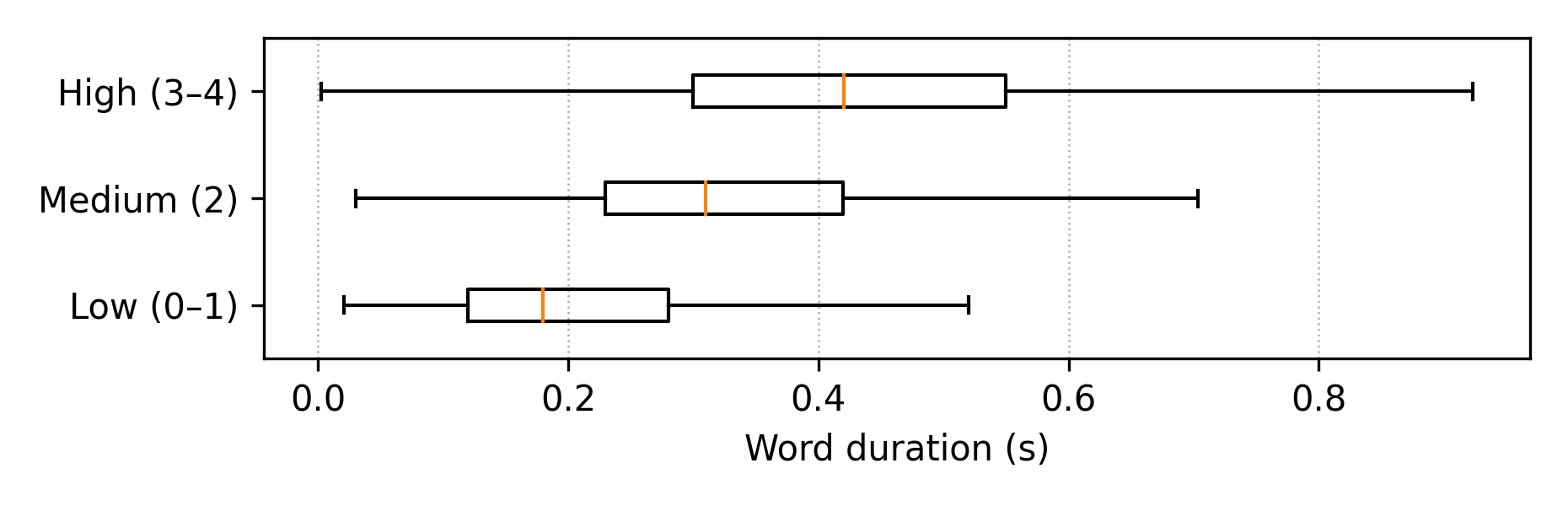}
\caption{Word duration by prosodic boundary strength (ToBI-style break index; backchannels excluded). Categories: low (0--1), medium (2), high (3--4).}
\label{fig:breakindex}
\end{figure}

If turn-final lengthening reflects boundary-related timing control, word duration should also track independently annotated prosodic boundary strength. We test this using ToBI-style break indices available in Switchboard and BU Radio, grouping indices into low (0--1), medium (2), and high (3--4) categories \cite{BeckmanAyersElam1997}. Fig.~\ref{fig:breakindex} shows a clear increase with boundary category: low-boundary words average $0.22$\,s, medium $0.34$\,s, and high $0.44$\,s (backchannels excluded), a $215$\,ms high-vs.-low difference (ratio $1.97\times$; $d=1.35$). This boundary-strength effect parallels the turn-position effect and supports the interpretation that turn-final lengthening is closely tied to prosodic boundary marking \cite{Wightman1992,TurkShattuck2007,White2014}.

\subsection{Cross-linguistic comparison suggests universal lengthening effect}
To reduce style confounds, we compare the two read-speech corpora (Table~\ref{tab:corpus}). Both show clear turn-final lengthening: BU Radio (English) $+201$\,ms ($d=1.08$) and Glissando (Spanish) $+254$\,ms ($d=1.47$). It proves to be robust in reading speech in both languages, with a stronger effect in Spanish in this comparison.

\section{Robustness Analyses}

To verify that the main turn-final lengthening effect is not driven by confounds, we perform three control analyses: (i) we isolate the contribution of backchannels, (ii) we stratify by word length, including a syllable-count stratification using corpus-provided syllable segmentation, and (iii) we examine variability across individual conversation sides.

\subsection{Backchannels}
Short listener responses can occur at turn boundaries but often follow different prosodic patterns. Table~\ref{tab:ablations} shows that backchannels alone show minimal lengthening ($\Delta=22$\,ms; $d=0.13$), and excluding them increases the effect ($d=1.22$ vs.\ $d=1.14$). The main effect is thus not driven by backchannels, whose inclusion only slightly attenuates it.

\subsection{Word length}
We stratify tokens by orthographic length to test if longer words drive the effect. Tab.~\ref{tab:ablations} shows that turn-final lengthening is present for short (1--4 chars: $\Delta=143$\,ms, $d=1.07$) and long (5+ chars: $\Delta=141$\,ms, $d=0.85$) words, with the standardized effect larger for shorter words, suggesting turn-final timing adjustments are not a by-product of word selection. Stratifying by syllable count using Glissando's word-level syllable segmentation confirms this pattern: monosyllabic ($d=3.08$), disyllabic ($d=1.42$), polysyllabic ($d=0.75$).

\begin{table}[t]
\centering
\caption{Robustness analyses of turn-position effect. Top: pooled across all corpora. Bottom: syllable-count stratification (Glissando only)}.
\label{tab:ablations}
\setlength{\tabcolsep}{4.2pt}
\begin{tabular}{lrrrr}
\toprule
Condition & $N_{TF}$ & $N_{MS}$ & $\Delta$ (ms) & $d$ \\
\midrule
Full dataset & 39{,}470 & 206{,}268 & 191 & 1.14 \\
Backchannels excluded & 35{,}947 & 195{,}266 & 203 & 1.22 \\
Backchannels only & 3{,}523 & 11{,}002 & 22 & 0.13 \\
Short words (1--4 chars) & 17{,}124 & 145{,}196 & 143 & 1.07 \\
Long words (5+ chars) & 22{,}346 & 61{,}072 & 141 & 0.85 \\
Very short (1--2 chars) & 3{,}011 & 57{,}914 & 105 & 0.85 \\
\midrule
Monosyllabic (1 syl) & 403 & 17{,}738 & 172 & 3.08 \\
Disyllabic (2 syl) & 2{,}081 & 8{,}044 & 118 & 1.42 \\
Polysyllabic (3+ syl) & 4{,}437 & 9{,}741 & 97 & 0.75 \\
\bottomrule
\end{tabular}
\end{table}

\subsection{Conversation-side variability}
\begin{figure}[t]
\centering
\includegraphics[width=0.95\linewidth]{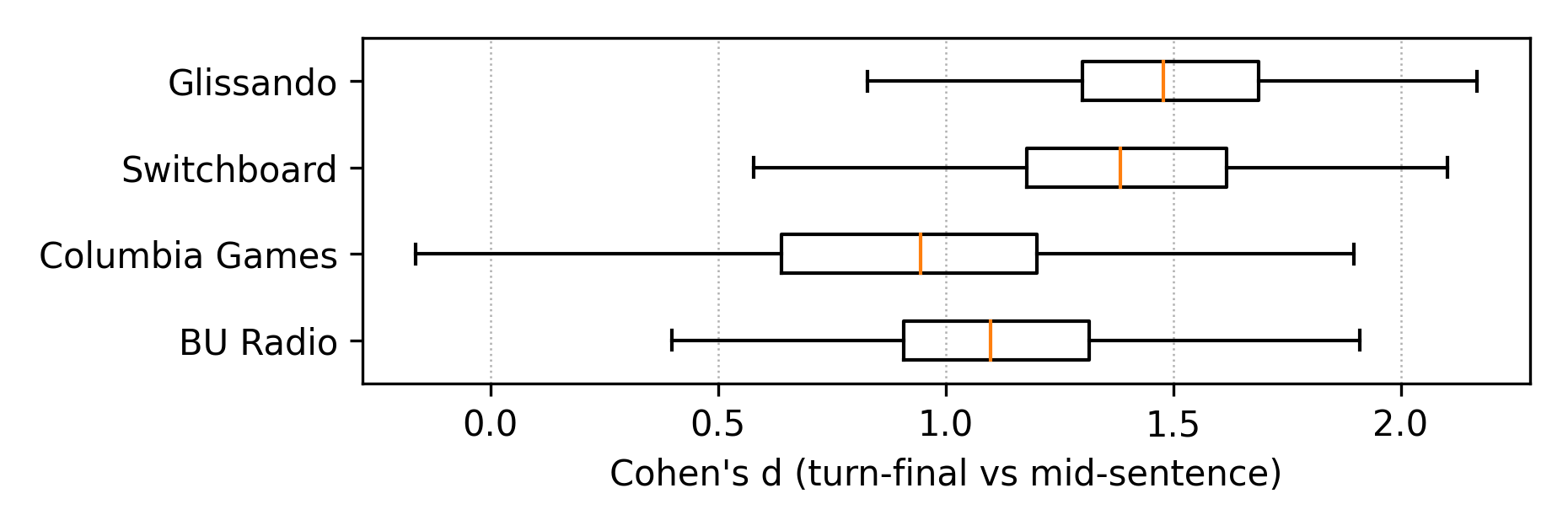}
\caption{Conversation-side variability in turn-final elongation (Cohen's $d$; backchannels excluded).}
\label{fig:unit_variation}
\end{figure}

Because ``speaker'' labels denote the channel/side of a recording rather than a persistent identity, we quantify variability at the level of \textbf{conversation side}\blue{(a single speaking channel within one recording), which does not need to correspond to a unique individual across recordings}, computing Cohen's $d$ per unit (requiring $\geq$10 tokens per condition). Figure~\ref{fig:unit_variation} shows strongly positive median effects in all corpora, with the largest effects in Glissando and Switchboard. Only a small fraction of units show non-positive effects (Columbia Games only), indicating the phenomenon is not driven by a subset of recordings.

\section{Discussion}
Across four corpora, turn-final words show significant lengthening relative to mid-sentence words. \blue{We frame these as position-dependent duration differences: they are equally consistent with lengthening at turn ends and with relative shortening of mid-utterance words, and our design does not adjudicate the direction of the effect.} The matched-word analysis demonstrates that the effect is not purely a lexical-selection artifact: the same word from the same speaker is reliably longer at turn endings. Break-index results align with a prosodic interpretation: stronger boundaries are associated with markedly longer word durations \cite{BeckmanAyersElam1997,Wightman1992,TurkShattuck2007}. Finally, syllable analyses support edge localization, consistent with accounts of boundary-adjacent timing control \cite{ByrdSaltzman2003} and word-edge effects \cite{WindmannSimkoWagner2015}.

Our pooled effect (203\,ms) exceeds read-speech estimates (40--100\,ms \cite{Oller1973,Klatt1975}), likely because turn boundaries are stronger prosodic boundaries and word-level measures accumulate across segments. The matched-word effect (80\,ms, $d=0.59$) is a more conservative, lexically controlled estimate.

``Turn-final'' is a mixed label, bundling prosodic boundary marking with interactional timing \cite{LevinsonTorreira2015,LocalWalker2012}. This harmonizes a large word-level difference with a more localized syllable-level pattern. These effects (${\sim}200$\,ms pooled, ${\sim}80$\,ms within-speaker) provide concrete targets for turn-taking models \cite{raux2009finite,skantze2021turn} and speech synthesis.

Why does the final syllable lengthen? Non-final syllables show a negligible turn-position contrast ($d=0.01$), whereas final syllables show a \blue{larger effect ($d=0.09$)}. This is consistent with Byrd and Saltzman's \cite{ByrdSaltzman2003} $\pi$-gesture model: a boundary-associated timing gesture slows articulation in its temporal neighborhood, with the strongest effects on the immediately adjacent segment. Our data support this localized account over a global-slowing interpretation.

\subsection{Limitations}
Corpus heterogeneity limits direct cross-corpus comparison, and we do not model speech rate or syntax in a unified regression. Our cross-linguistic comparison covers only English and Spanish (both stress-timed). Generalization to syllable-timed (e.g., French), mora-timed (e.g., Japanese), or tone languages (e.g., Mandarin) remains open. \blue{We also do not control for information structure (e.g., focus, givenness), which can modulate duration and may contribute to the differences we report. Finally, corpus-provided word boundaries may be less precise near turn ends, though the same-word controls, the boundary-strength gradient, and the within-word localization argue against annotation artifacts as the sole driver.}

\section{Conclusion}
Across four corpora spanning conversational, task-oriented, and read speech, turn-final words are $203$\,ms longer than mid-sentence words ($d=1.22$), with consistent effects across all corpora (Table~\ref{tab:corpus}).
% [CR] Removed a sentence here that restated the style-matched cross-linguistic result already given in Sec. 5.5: "In a style-matched cross-linguistic comparison, both BU Radio (English) and Glissando (Spanish) show clear lengthening."
Matched-word analyses confirm this is not a lexical-selection artifact ($d=0.59$), and syllable analyses localize the effect to the final syllable ($d=0.89$). These results support a localized, boundary-related account across the datasets and languages examined (English and Spanish), and provide quantitative constraints for models of prosody and turn-taking.\footnote{\blue{Supplementary materials: \url{https://mateocamara.com/word-lengthening/}}}
% [CR] Removed the closing future-work sentence to meet the 4-page content limit; this material is already covered by the Limitations subsection (cross-linguistic generalization, speech-rate normalization, mixed-effects models). Original:
% Future work includes typologically diverse languages, speech-rate normalization, mixed-effects models, and segmental analyses of syllable structure and stress.

\section{Acknowledgments}
This work was supported by the Ministry of Economy and Competitiveness of Spain under grant PID2021-128469OB-I00, and by the ``Ayuda Econ\'{o}mica Para Personal Investigador Postdoctoral 2025'' of Fundaci\'{o}n Santander. It was also supported by a grant from the MISTI MIT Global Experiences program.

\section{Generative AI Use Disclosure}
Generative AI tools have only been used for editing/polishing (not for writing any major part). The authors reviewed all the text.

\bibliographystyle{IEEEtran}
\bibliography{references}

\end{document}